\documentclass[useAMS,usenatbib,usegraphicx]{mn2e}
\usepackage{amsmath}

\newcommand\aj{AJ}
\newcommand\apj{{ApJ}}%
\newcommand\apjl{{ApJ}}%
\newcommand\apjs{{ApJS}}%
\newcommand\aap{A\&A}%
\newcommand\mnras{{MNRAS}}%
\newcommand\nat{{Nature}}%
\newcommand\physrep{{Phys.~Rep.}}%
\title[Physics of Low-Mass Galaxies]{The Effects of X-Ray and UV Background Radiation on the Low-Mass Slope of the Galaxy Mass Function}

\author[D.C. Hambrick et al.]{D.C. Hambrick$^{1}$\thanks{E-mail:
dclayh@astro.princeton.edu}, J.P. Ostriker$^{1}$,
  P.H. Johansson$^{2}$, and T. Naab$^{2}$\\
$^{1}$Princeton University Observatory, Princeton, NJ 08544\\
$^{2}$University Observatory Munich, Scheinerstr. 1, 81679 Munich, Germany}
\begin{document}
\maketitle

\begin{abstract}
Even though the dark-matter power spectrum in the absence of biasing
predicts a number density 
of  halos $n(M) \propto M^{-2}$ (i.e. a Schechter $\alpha$ value of $-2$)
at the low-mass end $(M<10^{10}M_{\odot})$, 
hydrodynamic simulations have typically produced values for stellar
systems in good
agreement with the observed value $\alpha\simeq -1$.  We explain
this with a simple physical argument and show that an efficient
external 
gas-heating mechanism (such as the UV background included in all
hydro codes) will produce a critical halo mass below which halos
cannot retain their gas and form stars.  We test this conclusion with
GADGET-2-based simulations using various UV backgrounds, and for the
first time  we also investigate the effect of an X-ray background.  We
show that at the present epoch $\alpha$ is depends
primarily on the mean gas temperature at the star-formation epoch
for low-mass systems ($z\la 3$): with no 
background we find $\alpha\simeq -1.5$, with UV only $\alpha\simeq
-1.0$, and with UV and X-rays $\alpha\simeq -0.75$.  We find the
critical final halo mass for star formation to be $\sim 4\times 10^8
M_{\odot}$ with a UV background and $\sim 7\times 10^8
M_{\odot}$ with UV and X-rays.    
\end{abstract}


\section{Introduction}

The mass spectrum of dark-matter (DM) halos collapsing from an initial
gaussian-perturbed density field has long been an object of study; the
formalism most widely used today is that of \citet{ps74}, who derived
a number density of the form 
\begin{equation}
n(M)dM=M^{-2}\bar{\rho}\sqrt{2/\pi}\frac{\delta_c}{\sigma}\frac{d\ln\sigma^{-1}}{d\ln
  M}e^{-\delta_c^2/2\sigma^2}dM
\end{equation}
where $n(M)dM$ is the number of halos in the mass range $M$ to $M+dM$,
$\bar{\rho}$ is the mean mass density, $\delta_c$ is the critical
(linear) density for collapse, and $\sigma$ is the rms perturbation.
If we consider small scales, where $\sigma$ is both $\gg\delta_c$ and
approximately independent of mass, then the Press-Schechter form
reduces to 
\begin{equation}
n(M)dM \propto M^{-2} dM
\end{equation}
\citep{cgo01}.  

  The luminosity function of galaxies is commonly
 parametrized in the Schechter form \citep{s76}, 
\begin{equation}
n(L)dL\propto (L/L_*)^{\alpha} e^{-L/L_*}dL;
\end{equation}
i.e., a power-law of slope $\alpha$ at the low-luminosity end with an
 exponential cutoff at a characteristic luminosity $L_*$.  If star
 formation traced the underlying dark matter distribution exactly
 (i.e. a constant mass-to-light ratio), we
 would expect to observe $\alpha\simeq -2$ at the faint end.  However,
 galaxy surveys 
 such as the SDSS have consistently shown a 
 different slope, in the range of $-1.2\la\alpha\la -0.9$
 (see \citealt{blant01}, Table 2 of \citealt{ac05}, and
 \citealt{pg08}).  The extent to 
 which galaxies or dark-matter halos fail to follow  
the clustering distribution of linear perturbation theory is known
 generally as ``bias'', and has been 
 an active area of phenomenological and theoretical research
 (see, e.g., the introduction to \citealt{bens00} and \S6 of
 \citealt{cs02} for an overview): most relevant here is the relative bias
 between galaxies and DM halos, i.e. the extent to which galaxies fail
 to follow the underlying DM.  At the low-mass end, the
 discrepancy has come to be called the ``missing satellites problem''
 \citep{kkvp99}, since  na\"\i ve Press-Schechter theory predicts
 $\sim 300$ satellite galaxies for the Local Group, while only $\sim 40$ are
 observed \citep{sg07}---although this number continues to increase
 as data from e.g. SDSS are analyzed.  

One might expect that this relative bias would be a concern for those
performing galaxy simulations, and indeed DM-only simulations which
assign galaxies to DM halos after the fact \citep[e.g., the Millennium
  Run of][]{spring05} do use an ad-hoc prescription to reduce the
baryonic mass in small halos, or overpredict small halos
\citep{dlb07}, although agreement is improving as semianalytic
models improve in sophistication \citep{guo10b}.  However,
hydrodynamic simulations seem to 
require no such prescription: they most often report a faint-end mass
spectrum slope in the vicinity of -0.9 -- -1.2 \citep{cgo01,nchos04,ac05}.
While there are some exceptions (\citealt{opp09} report $\alpha=-1.45$
for their best-fit model; \citealt{crain09} find $\alpha=-1.96$), it would
appear this is a result which is difficult to get wrong.

\citet{ds86} were the first to propose a mechanism of gas being heated
(by local supernovae) and driven out of low-mass halos. \citet{efst92}
introduced the idea of the ionizing background radiation being the
relevant mechanism, and \citet{qke96} used a smooth particle hydrodynamic
(SPH) 
simulation to show that a photoionizing background does strongly
inhibit the cooling and infall of gas into halos of virial mass below
$4\times 10^9 M_{\odot}$.  \citet{tw96} did the same with a 1D
Lagrangian code. \citet{ns97} showed that the UV background reduced
the cooled gas accreted to disk galaxies by half, with late-accreting
gas preferentially affected.  
\citet{gnedin00} used simulations of reionization to determine
that the mass scale at which gas is 
stripped by photoionization from the background radiation corresponds
to the ``filtering'' length scale, the scale at which baryonic matter
is smoothed compared to the underlying dark matter in linear
perturbation theory; however \citet{ogt08} using GADGET-2 found a somewhat
smaller critical mass of $10^{10} M_{\odot}$ (similar values
  were obtained by \citealt{hoeft06} and \citealt{crain07}) and, significantly,
reported that their SPH result was well approximated by a simple
prescription comparing the gas temperature to the virial temperature
of the halo.   \citet{bens02} and
\citet{som02} made 
semianalytic calculations of galaxy evolution and likewise found that
galaxies fainter than $L_*$ 
have their star formation suppressed or ``squelched'' by a
photoionizing background. 
Meanwhile, the role of supernova feedback in driving gas from
small halos and thereby suppressing star formation has continued to be
studied \citep{yep97,kay02,scan08}.
 \citet{ps10} have even performed a
simulation showing that the UV output of local stars can affect the
SFR.  However, while these processes are certainly present and
important, our focus here is on the ionizing background, which we
will show to be both necessary and sufficient to reproduce the observed
Schechter $\alpha$.

It should be noted that despite the terminology of ``critical'' or
``cutoff'' masses, a sharp cutoff in galaxy masses is not consistent with the
observations: there are low-mass (dwarf) galaxies, merely fewer than
 expected.  Several mechanisms have been proposed to explain
this. \citet{sd03} proposed that some intermediate-mass halos
($v_{\text{circ}}\sim 25-35$ km/s, or $M \sim 10^{10} M_{\odot}$) have
star formation at early times, then lose their remaining gas at the
time of reionization and become ``red and dead'' dwarfs at the
present.  \citet{kgk04} suggested that present dwarf satellites such
as those around the Milky Way were originally larger but were reduced
to their present size by tidal stripping.  Such mechanisms may be
needed to explain  the very low luminosity ($\ga 10^3
L_{\odot}$) Milky Way 
satellites discovered in recent years \citep{will05,belo06,belo07}.

Two recent studies are of particular interest. \citet{sawl09} studied the
evolution of dwarf galaxies in isolation and found that both feedback
and a UV background are necessary to reproduce the observed properties
of the Local Group dwarf spheroidals.  In particular, they saw that
the UV background efficiently suppresses star formation after the
epoch of reionization for systems below $10^9 M_{\odot}$.  Similarly,
\citet{of09} simulate a Milky Way-like galaxy and its satellites and
find a threshold circular speed of 12 km/s ( $\sim 8\times 10^8
M_{\odot}$) for halos which can form stars.  

We also note that even though it has been known at least since
\citet{me99} that including the X-ray background has a substantial
effect on the 
tempterature of the IGM (and hence, as we will see, on star
formation), few simulations have incorporated this component.
\citet{ric08} studied its effect on dwarf galaxies and found a
critical baryon retention mass, but only looked at
$z>8$.  We remedy this deficit by including an X-ray background in
some of our simulations to study its effect on small galaxies from
moderate redshift to the present.  

In a previous paper \defcitealias{self09}{Paper I}\citep[][hereafter
  Paper I]{self09}, we explored the
effect of the ionizing background on the stellar and gas properties of
large elliptical galaxies.  Here, we use a similar set of UV and
  X-ray
  backgrounds to examine their effect on the low end of the galaxy mass
  spectrum, namely the satellites of those ellipticals.  Section 2 presents
  a simple physical argument for why any 
  efficient gas-heating
  mechanism will yield a flattening of the low-mass slope.  Section 3
  presents the details of the simulations we performed to test this
  argument; section \ref{sect:results} gives the results of the same.
  Section \ref{sect:disc} is discussion and   conclusions.

\section{Physical Argument}\label{sect:theory}
Our argument, which dates back to \citet{rees86} and \citet{efst92},
compares the escape velocity of gas in a virialized halo 
to the sound speed of that gas.  The escape velocity for a halo of
mass $M$ and radius $r$ is given by
\begin{equation}
v_{\text{esc}}=\sqrt{\frac{2GM}{r}}.
\end{equation}

We relate $M$ and $r$ by fixing the density at the standard
``virialized'' value, 
\begin{equation*}
\rho=200 \bar{\rho}
\end{equation*}

where $\bar{\rho}$ is the mean density of the universe at a given
epoch (we calculate the result for $z=0$, and assume $\Omega_M=0.3$,
as in our simulations).  Then
the escape velocity becomes 

\begin{eqnarray}
v_{\text{esc}}&=&\sqrt{2G \left(\frac{800 \pi}{3} \bar{\rho}M^2\right)^{1/3}}\\\nonumber
&=&36.0 \left(\frac{M}{10^{10}M_{\odot}}\right)^{1/3} \text{km/s}.
\end{eqnarray}

The sound speed, meanwhile, is given for a gas of temperature $T$ and
mean particle mass $m$ by 

\begin{eqnarray}
v_{c}&=&\sqrt{\frac{\gamma k T}{m}}\\\nonumber
&=&21.7 \left(\frac{T}{10^{5} \text{K}}\right)^{1/2} \text{km/s},
\end{eqnarray}

where we assume the adiabatic index $\gamma = 5/3$ for non-relativistic
gas, and for $m$ we assume primordial abundances of H and He.  By
equating these speeds, we have a formula for a critical mass: 
\begin{equation}
M_{\text{crit}}=2.19\times 10^9
\left(\frac{T}{10^{5} \text{K}}\right)^{3/2} M_{\odot}
\end{equation}
At 
a given gas temperature, halos below this critical mass
 will have gas 
thermal speeds greater than their escape velocities.  Hence they will
not be able 
to hold on to their gas, and therefore be unable to form stars.  For
a temperature of $2\times 10^4$ K, corresponding to the peak of the
H cooling curve,  $M_{\text{crit}} = 2 \times 10^8 M_{\odot}$.
Therefore we expect halos with less than this total mass to have very
little star formation, and at late times, very little baryonic mass of
any kind, assuming there is a mechanism to heat the gas to $2\times
10^4$ K by the epoch where star formation in these halos would be
significant.   Heating above this temperature will further increase
the cutoff mass.   Moreover, since larger systems are formed by the 
hierarchical assembly of smaller ones, we expect this baryon deficit
to creep to larger-mass halos as time passes, effectively increasing
the critical mass at the present by some factor $f \ga 1$.

\section{Simulations}
To verify the physical argument presented above, we analyze several
high-resolution simulations that are performed with various ionizing
(UV and X-ray) backgrounds.  These simulations were performed with a
slightly updated version of the GADGET-2-based code used in
\citetalias{self09}; refer to that work for 
additional details.  In brief, 
the code is based on GADGET-2, and the halos
are selected from a (50 Mpc)$^3$ box with cosmological ($\Lambda$CDM)
initial conditions, and resimulated in a (1 Mpc)$^3$ box centered on
each halo, with the central (0.5 Mpc)$^3$ using high-resolution DM
and gas/star particles \citep[for more details see][]{naab07,jno09}.  Since we specifically select high-density
regions containing massive galaxies, our results are not necessarily
applicable to other environments.  To ameliorate the effects of
cosmic variance, we choose seven galaxies (that is,
seven initial-conditions (IC) boxes), with
final virial 
masses in the range $1-2\times 10^{12} M_{\odot}$.  
The simulations do not include optical
depth effects, in particular the self-shielding of dense star-forming
regions from the UV background.  They do (differently from
\citetalias{self09}) include 
feedback from supernovae though not AGN, following the two-phase
\citet{sh03} model, and a simple prescription for
metal-line cooling using cooling rates calculated by Cloudy \citep[v07.02,
last described in][]{cloudy}, which presumes photoionization and
collisional equilibrium for the gas and metal atoms, and assumes $0.1$
solar metallicity.  Star 
formation is 
performed at a fixed density threshold calculated by the code such
that $\sim 90\%$ of the gas above the threshold will be in cold phase:
this threshold is $\rho_{\text{crit}}=1.6\times
  10^{-25}$~g~cm$^{-3}$ or $n_{H,\text{crit}}=0.07$~cm$^{-3}$.  Star
  particles have half the mass of original gas particles; thus each
  gas particle can ultimately turn into two star particles. 
  As in \citetalias{self09}, all simulations
were performed with initially $100^3$ each of SPH (i.e. baryon)
particles and DM 
particles, with a 
gravitational softening length of 
0.25 kpc for the gas and star particles and twice that for the dark
matter particles.  Gas and star particles have masses in the range
$4 - 7 \times 10^5 M_{\odot}$, depending on the size of the box; 
the assumed cosmology is
$(\Omega_M,\Omega_{\Lambda},\Omega_b/\Omega_M,\sigma_8,h)=(0.3,0.7,0.2,0.86,0.65)$.

We run these simulations using a variety of ionizing backgrounds:
the same set which was used in \citetalias{self09}, with two
additions.  The first, which we call
No UV, has as the name suggests no ionizing radiation at all.  The
next three backgrounds are 
Old UV, an updated version of \citet{hm96}; New UV, which has the same
spectrum as Old UV but an intensity (as measured by e.g. the H
photoionization rate) which declines as $(1+z)^{-1}$ above the peak at
$z\simeq 2.2$; and FG UV, the background constructed in \citet{fg09}.
The final two backgrounds add an X-ray component, which has a spectrum
taken from \citet{sos04} and a 
normalization from \citet{gilli07}.  The New UV+X model takes the New
UV model and adds this X-ray component with the same redshift
dependence, while FG UV+X adds this component to the FG UV background but
with the redshift dependence of Old UV \citep[i.e.][]{hm96} to more
closely replicate the quasar background that the X-ray component is
meant to model.  The quasar background peaks at $z\sim 2$ and becomes
negligible for $z\ga 3$ \citep{hop07}.

\section{Simulation results}\label{sect:results}
\subsection{Results at $z=0$}
Having performed the simulations, we identify virialized systems in
both the star 
and total (stars plus gas plus DM)  particle populations using AHF,
the Amiga Halo Finder 
\citep{kk09}.  AHF uses a density grid-tree method followed by the
removal of unbound particles to find halos and subhalos, and returns
the virial mass of each.  The virial mass is defined as the total mass
of all (relevant) particles inside the virial radius, which in turn is
defined as the radius where the calculated density profile exceeds a
threshold multiple of the critical density determined by cosmology and
redshift ($\sim 200$); for subhalos the virial radius is instead the
largest radius of any particle which is gravitationally bound to the
halo.  See the AHF 
documentation for details. 

\begin{figure}
\includegraphics[width=\linewidth]{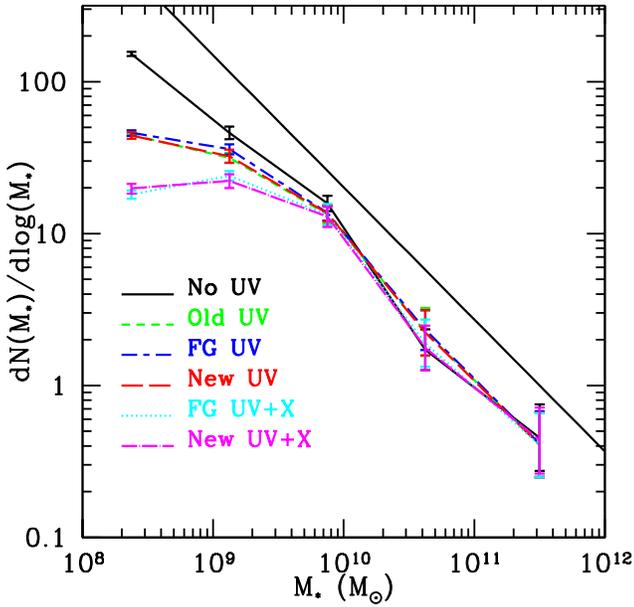}
\caption{The average number of stellar systems (galaxies)
  per decade 
  in stellar mass 
  identified by AHF with mass 
  in given range, at $z=0$. The error bars represent $1\sigma$
  statistical errors among the 7 initial conditions.  Also included is
  an $M^{-2}$ power law for comparison.  The number of $<10^9$
  M$_{\odot}$  galaxies 
  decreases markedly as one adds first UV and then X-ray radiation.} 
\label{fig:fracts}
\end{figure}

 Figure~\ref{fig:fracts} shows the mass
distribution of 
stellar systems (identified using star particles only).  AHF can
identify systems as small as $1.5\times 10^7 M_{\odot}$, or about 40
star particles, but we adopt $10^8 M_{\odot}$ as a lower mass limit to
ensure completeness.  The mean 
number of systems (averaged over the seven 
ICs) in the lowest mass decade $(10^8 -
10^9 M_{\odot})$ declines from $131$ with No UV to $\sim 43$ with the three
UV models  and $\sim 20$ for the two UV plus X-ray cases, 
differences of $36\sigma$ and $14\sigma$ respectively, while
the number in the two highest mass bins $(>10^{10} M_{\odot})$ remains
nearly constant.   As an aside, we expect these highest-mass bins will be high
compared to the 
global spectrum because we specifically select boxes which contain a
massive galaxy; furthermore, systems of this virial mass
($M_{\text{Halo}}\approx 
2\times 10^{12} M_{\odot}$) have stellar masses which are very vulnerable to
suppression  by AGN
feedback \citep[e.g.][]{mcc10}, not included here.    

We also see that the UV-only models (Old~UV, FG~UV and New~UV)
are essentially identical to each other in mass spectrum.  The two
models with X-rays (New~UV+X and FG~UV+X) are also identical to each
other, though of course quite different from the models without X-rays.
This suggests that the details of
the ionizing background above $z=3$, including the epoch of
reionization, are relatively unimportant in determining the mass
spectrum. 

Using the Bayesian inference method of \citet{kfv08}, we estimate
the Schechter parameters $\alpha$ and $M_*$; the results for $\alpha$
are presented 
in Table~\ref{tab:sch}. This method has
the advantage of not requiring any binning of the data.  Values for
$M_*$ were roughly constant for 
the various background models in the
range $10^{9.5-10} M_\odot$; also note that we use stellar mass as a
proxy for luminosity without considering, e.g., age.    Consistent with
the appearance of  
 Figure~\ref{fig:fracts}, $\alpha$ increases dramatically from $-1.5$
 for the no-background case to roughly $-1$ for the UV-only models and
 $-0.75$ for the models with X-rays.  Thus we see that the addition of
 a simple UV ionizing background was the only change necessary (with
 feedback; see below) to move
 the low end of
 the stellar mass spectrum into consistency with observations, while
 an additional X-ray background (which most simulations historically
 have not used) seems to make the slope \emph{too} flat.

\begin{table}

\caption{Schechter $\alpha$ values and maximum ``barren'' halo
  masses with various backgrounds.
\label{tab:sch}}
\begin{center}\begin{tabular}{lccc}
\hline
Name & $\alpha$ & $\log f M_{\text{crit}}$ & $\epsilon_{\star}$\\
\hline
No UV & $-1.52 \pm 0.06$ & $8.19\pm 0.06$ & $0.47\pm 0.02$\\
Old UV & $-1.01 \pm 0.04$ & $8.55\pm 0.08$ & $0.37\pm 0.02$\\
FG UV & $-1.03 \pm 0.06$ & $8.50\pm 0.08$ & $0.37\pm 0.02$\\
New UV & $-1.08\pm0.05$ & $8.63\pm 0.05$ & $0.37\pm 0.02$\\
FG UV+X & $-0.73 \pm 0.04$ & $8.95\pm 0.07$ & $0.32\pm 0.02$\\
New UV+X & $-0.78\pm0.08$ & $8.78\pm 0.04$ & $0.32\pm 0.02$ \\
\hline 
No UV-F & $-1.64 \pm 0.07$ & $8.06\pm 0.06$  & $0.48\pm 0.04$\\
FG UV-F & $-1.29 \pm 0.06$ & $8.22\pm 0.07$  & $0.38\pm 0.04$\\
FG UV+X-F & $-1.16 \pm 0.11$ & $8.22\pm 0.09$  & $0.33\pm 0.04$\\
\hline
\end{tabular}\end{center}

\medskip
The Schechter-$\alpha$ values for the star
  particles, maximum ``barren''
  halo masses, and overall baryon-conversion efficiency with various
  backgrounds.   
  Adding UV and X-ray heating steadily increases
  $\alpha$: models with UV only are consistent with observations.
 Similarly, the critical mass for baryon retention 
  increases, and the baryon-conversion efficiency decreases, when the UV
  background and X-rays are added.  The ``-F'' rows represent
  simulations with no SN feedback; see text for details.

\end{table}

To directly compare these results with the mechanism presented in
\S\ref{sect:theory}, we estimate $f M_{\text{crit}}$ (the effective
star-formation cutoff mass) by the following
method.  We take the mass spectrum of halos identified by AHF from the
total population of particles (stars, gas and DM), and find for each
the total stellar mass in the halo.  Then starting at the low-mass end of
the spectrum we consider an 11-point moving average of the ratio of
total stellar mass ($M_\star$) to total overall virial mass ($M_\text{Halo}$)
among the halos.  (The number 
11 was chosen to balance between mass resolution and the inherent
scatter in stellar mass to total mass ratio among small halos, but
even using a 41-point average does not significantly change the
results.)  We move
upward through the mass spectrum until this ratio exceeds some fixed
value, which we choose to be half of the global average 
($\sim\! 0.16$), which is the same definition as 
\citet{gnedin00} and \citet{ogt08}.  The particular choice of critical
mass is not very important, as the baryon-conversion efficiency is a
steep function of halo mass around the critical point \citep[see
  Figure 2 of][]{ogt08}---even if we choose a threshold of one-tenth
the global average,
$f M_{\text{crit}}$ falls by only $\sim\!0.2$ dex.  Thus 
$f M_{\text{crit}}$ represents the largest total mass at which systems
have severely restricted star formation.  Mathematically:
\begin{equation}
f M_{\text{crit}}= M_{\text{Halo}}(N)\ni \frac{1}{2}\frac{\displaystyle\sum_{i}M_{\star}(i)}{\displaystyle\sum_{i}M_\text{Halo}(i)} = \frac{\displaystyle\sum_{i=N-5}^{N+5}M_{\star}(i)}{\displaystyle\sum_{i=N-5}^{N+5}M_\text{Halo}(i)}
\end{equation}

  The results are presented
in the third column of Table~\ref{tab:sch}. 
We find the critical mass in the No UV case
is $1.5\times 10^8$~M$_{\odot}$, but when a UV background is added this
increases by a factor of more than 2 to
  $\sim\! 3.6\times 10^8$~M$_{\odot}$,
 and with X-rays by another factor of 2 to
 $\sim\! 7.3\times 10^8$~M$_{\odot}$, in good agreement with
 \citet{of09}, who also looked at satellite galaxies. They used a
  background with UV but not X-rays, but they did include supernova 
 winds, which should act in a similar way as X-rays to blow out gas
 from small halos; 
 thus we would expect our UV+X critical mass to be the most similar to
 theirs, which is what we find.  
 \citet{ogt08}, meanwhile, report a significantly higher 
 value of $\sim 10^{10}$~M$_{\odot}$ using only isolated halos; see the
 discussion in the next paragraph.
We find (see \S4.2) that No UV has a gas temperature at the
star-forming epoch of $2\times 10^4$~K, so comparing that cutoff mass to
the value of  $2\times 10^8$~M$_{\odot}$ 
derived in \S\ref{sect:theory}, we find that the dimensionless
hierarchical-structure correction factor $f\approx 0.75$, actually less
than unity.  This seems to support the idea of \citet{kgk04} that small
satellites 
at late times were originally larger, and indicates that systems above
the critical mass are able to accrete sufficient gas to overcome the
deficit of accreted stars.

Finally, we revisit the finding in \citetalias{self09} that overall
star formation is suppressed with the ionizing background.  To compare
across the various initial conditions, we calculate the baryon-conversion
efficiency, 
$\epsilon_{\star}$, defined as  
\begin{equation*}
\epsilon_{\star}\equiv \frac{M_{\star}}{M_b}
\end{equation*}
where $M_{\star}$ and $M_b$ are the total stellar and baryonic (gas
plus stars) masses, respectively, in the high-resolution simulation
box. The results are in the last column of Table~\ref{tab:sch},
These 
efficiencies are higher by a factor of roughly four than the observed
values, recently 
measured by \citet{guo10} to be $\sim\! 20\%$ (or $\sim\!
4\%$ as a fraction of total mass rather than baryons) for halos of
mass $\sim6\times 10^{11}$~M$_{\odot}$, and less for other halo
masses. Similar results were derived by \citet{most10} (who found that
efficiency is linear in mass at low masses) and \citet{tg10}.  

\begin{figure}
\includegraphics[width=\linewidth]{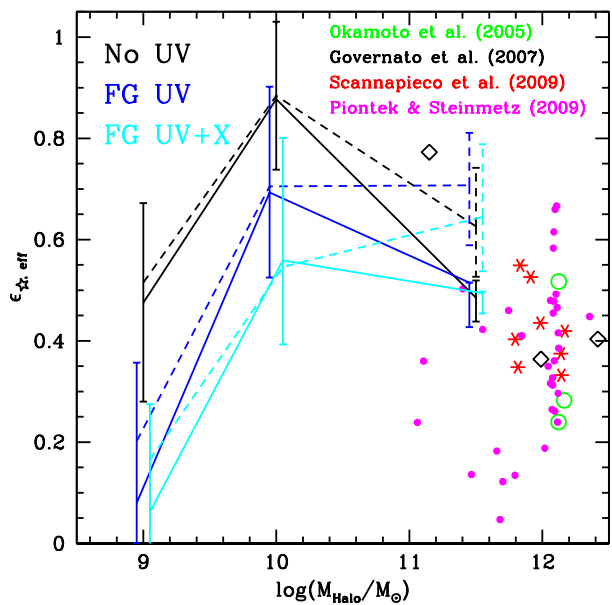}
\caption{
The baryon-conversion efficiency as function of (total, viral) halo
mass. The scatter 
points are previous 
simulations.  Our results, the connected error bars,  are
the median baryon-conversion efficiencies
for 1 decade wide bins in halo mass
across all 7 ICs; the error bars are the inter-quartile ranges, and
there is a small horizontal offset for clarity.   The solid lines are the
models with SN feedback and the dashed lines are the models without
it (which use 3 ICs rather than 7).   The
addition of the ionizing backgrounds reduces star formation by a
factor of $\la 5$, depending on the mass bin.
 } 
\label{fig:guo}
\end{figure}

Figure~\ref{fig:guo} shows the baryon-conversion efficiency as a
function of halo mass, using a different definition:
\begin{equation}
\epsilon_{\star,\text{eff}}\equiv
\frac{M_{\star,c}}{M_{b,\text{implied}}}= \frac{\Omega_M
  M_{\star,c}}{\Omega_b M_\text{Halo}}= 5 \frac{M_{\star,c}}{M_\text{Halo}},
\end{equation}
where $M_{\star,c}$ is the stellar mass within $0.1$ virial radii of
the halo center, similar to \citet{sal10}, and the ``implied''
baryonic mass
$M_{b,\text{implied}}\equiv M_\text{Halo}(\Omega_b/\Omega_M)$ is
simply the mass of baryons associated with the halo well before reionization.
Figure~\ref{fig:guo} is analogous to Figure~5 of \citet{guo10},
although the results are not directly comparable (see below).   There
is a factor of $1.5-5$ reduction in  
star formation with the addition of the
ionizing backgrounds, and that 
the effect is stronger in lower mass halos. However our results are
still too high compared to the SDSS results
by a factor of $\sim 2$ in the highest-mass bin, and more in the
lower ones.   
We ascribe this discrepancy first to fact that we are not looking at any
field galaxies, but only galaxies in a region specifically selected to
have an overdensity and a massive central galaxy. This is clearly indicated 
 in the figure, since SF efficiency actually increases
with decreasing halo mass from $10^{11.5} M_{\odot}$ to  $10^{10}
M_{\odot}$: a satellite galaxy will naturally have a smaller virial
radius than a field galaxy, and hence a smaller DM mass will be found by the
halo finder, not to mention the enhanced star formation from the
denser environment \citep{ogt08}.  
The lack of SN-driven 
winds in our 
simulations, which would otherwise eject gas from low-mass galaxies,
also makes a significant contribution;  
the relation between mass-loss in winds and the SFR has been verified
observationally \citep{rvs05} and theoretically \citep{od05}.
We also note that the
other recent simulations shown in
Figure~\ref{fig:guo}---\citet{oka05}, \citet{gov07}, \citet{scan09}, 
and \citet{ps09}---seem to have too much star formation by factors of
a few as well, and furthermore we are primarily interested in the relative
differences between the various background models: thus
an overall excess of star formation can be safely disregarded.
As
with $\alpha$ and $f M_{\text{crit}}$, there is no significant
difference between the three UV-only models, or between the two UV plus
X-ray models.

To explore the effect of the SN feedback
on our results, we ran three sets of
simulations (the A, C, and E ICs) using the No UV, FG UV and FG UV+X
backgrounds with no feedback.  The primary results are given in the
last three rows of Table~\ref{tab:sch}.  The Schechter $\alpha$ values
are somewhat steeper in each case compared to the runs with feedback,
which we expect since feedback gives a direct energy injection at the
site of star formation,
but the differential effect of the background is much the same,
although not quite as strong.  In particular, the FG~UV+X-F model
produces an $\alpha$ within the observed range, showing that SN
feedback, while certainly present and important, is not necessary to
produce the correct spectrum of low-mass galaxies.  A
similar result is obtained for 
$fM_{\text{crit}}$, which 
increases by roughly 0.2 dex from
no~UV to FG UV and changes negligibly when X-rays are added.  It
therefore appears that the 
effects of the ionizing background and of SN feedback amplify one
another in a superlinear fashion, as suggested by \citet{ps08} and
\citet{sawl09}.  This is also indicated by Figure~2, where the dashed
lines correspond to the no-feedback models: in the lowest-mass bin,
adding feedback causes a greater relative reduction in baryon-conversion
efficiency when an ionizing background is present. 

To make a detailed comparison with \citet{sawl09}, we calculate the
baryon-conversions efficiencies for a 0.2-dex-wide bin around their
fiducial halo mass of $7\times 10^8 M_{\odot}$.  For their runs with
UV background only, SN feedback only, and both, \citet{sawl09} give
efficiencies of 0.77, 0.05, and 0.03, respectively.  Here, however,
the respective models FG~UV-F, No~UV and FG~UV give median values of,
respectively, 
0.16, 0.47, and 0.04.  So for \citet{sawl09}, the feedback is by far
the most important contributor, whereas here the background is
somewhat more important.  The stronger effect of the background in
this work seems likely to be the effect of
timing: \citet{sawl09} turn their background on at $z=6$, by which
time their galaxy has already created $20-30\%$ of its final stellar
mass; in our simulations, however, galaxies of this final size have no
significant star formation until $z\sim 3$ (see the next subsection).
Therefore our galaxies show no significant change in baryon conversion
between Old~UV, which turns on at $z=6$, and FG~UV, which turns on at
$z=10$, but we speculate that if the \citet{sawl09} runs were to be
repeated with FG~UV, the background would have a more significant
effect.  The relatively weak effect of feedback for our models, on the
other hand, is likely the effect of environment: the relatively dense
environment and corresponding IGM pressure on our small halos means
that they do not have 
enough supernova energy to unbind their gas.  

 The physical
mechanisms governing
the interaction between feedback and the background radiation are
beyond the scope of this paper, but our 
results are consistent with a picture where feedback moves gas
from the centers of halos to the outer regions, where the background
then provides enough energy to unbind the gas altogether.  Thus there
is no change in the $fM_{\text{crit}}$ when X-rays are added in the
absence of feedback because the dense central gas has too short a
cooling time to be affected (although the smallest halos with the lowest
central densities can still be suppressed, increasing $\alpha$),
whereas if it were first pushed out 
somewhat by feedback the density would be lower, the cooling time
longer and X-rays could have an effect.  Conversely, with feedback and
no background (the No~UV case), the gas gets pushed outward, but not
enough to unbind, so it falls back in and forms more stars.  

\subsection{Evolution with Time}

\begin{figure}
\includegraphics[width=\linewidth]{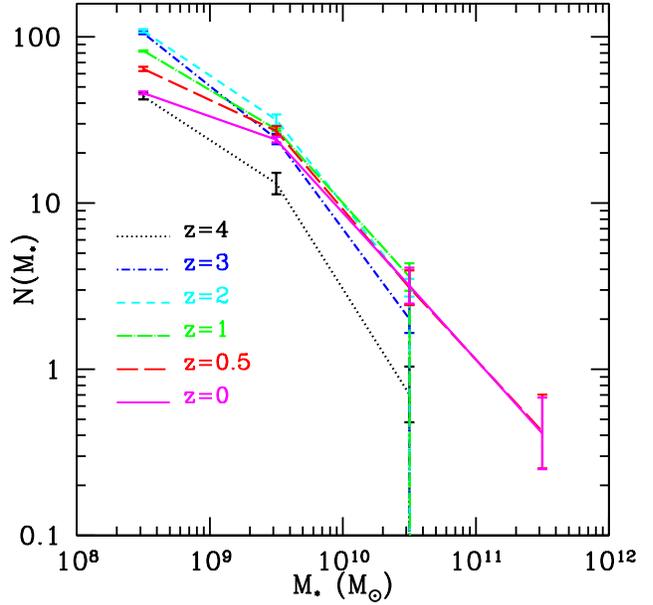}
\caption{The mass spectrum (average number of systems per decade of mass) of stellar
  systems with the FG UV background
  at various  epochs. The number of small galaxies peaks at $2\la z
  \la 3$.} 
\label{fig:nhist}
\end{figure}

To investigate the origins of the discrepancies in low-mass systems, we
examine the simulations at higher redshifts.  Figure~\ref{fig:nhist}
shows the stellar mass spectrum of the FG UV simulations from $z=4$ to
the present.  The number of low-mass systems peaks at $z\approx 2-3$ and then
declines as small systems merge into larger ones.  This peak is
somewhat later than the overall star-formation peak, found
in \citetalias{self09} to be at $z\approx 4$, which is precisely the
``downsizing'' effect first reported by \citet{cow96} \citep[see][for
  more recent observational results]{zheng07}.  This effect
explains why our three UV models show such similar results here: at
$z=4$ the Old UV and FG 
UV models have not finished HeII reionization, and Old UV has $\sim 3$
times lower ionization rate than the other two models, while for $z\leq3$ the
three models are nearly identical in ionization rate and have all fully
reionized \citepalias[see Figure 1 of][]{self09}.  Thus at the epoch when the smallest halos are forming
stars, the gas temperatures in the three models are much more similar
than when stars in more dense regions formed \citepalias[gas
  temperatures are plotted in Figure 2
  of][]{self09}. On the other hand, the models with X-rays have higher
gas temperatures than the UV-only models for all $z<4$, which is
reflected in the increased suppression of small systems. 

Table~\ref{tab:hist} compares the total number of identified (stellar)
systems in the whole box for the No UV and FG UV
cases, as well as the baryon conversion efficiency
 $\epsilon_{\star,\text{eff}}$, 
defined in Equation~9, for two
bins in halo mass ($10^{8.5}<(M_\text{Halo}/M_{\odot})< 10^{9.5}$ and 
$10^{10.5}<(M_\text{Halo}/M_{\odot})$).  Although 
No UV has $\sim 3$ times the number of systems  as FG UV, they both
show the same time dependence: the 
number of systems reaches a peak at
$2\la z \la 3$, followed by mergers reducing the number.  The stellar
mass fraction for high-mass halos is nearly the same for the two
models, while for low-mass
halos the difference is a factor of $2-6$.
Interestingly, in the FGUV case the low-mass stellar mass fraction declines
somewhat from $z=2$ to 0, presumably because closer satellites of the
central galaxy, which had been able to form more stars than the
average because of their denser environment, are finally accreted and
lost (while with No UV all satellites can form many stars).

\begin{table}
\caption{Stellar system statistics over time
\label{tab:hist}}
\begin{center}
\begin{tabular}{lcccccc}
\hline
  &\multicolumn{4}{c}{$\epsilon_{\star,\text{eff}}$} &
 \multicolumn{2}{c}{$N$} \\
&\multicolumn{2}{c}{low-mass} &
  \multicolumn{2}{c}{high-mass}  &  &\\
 z & No UV & FG UV & No UV & FG UV & No UV & FG UV\\
\hline
4 & .111 & .052 & .546 & .569 &153 & 63\\
3 & .266 & .140 & .610 & .618 & 425 & 159\\
2.5 &.325 &.180 &.640 & .650 & 503 & 174\\ 
2 & .400 & .220 & .646 & .637 & 567 & 169\\
1 & .467 & .165 & .728 & .730 & 526 & 134\\
0.5 & .475 & .118 & .739 & .792 & 449 & 109\\
0.0 & .475 & .080 & .787 & .776 & 328 & 82 \\
\hline
2.5 &.173 & .353 & .697 &.712 & 259 & 443 \\
0.5 & --- & .010  & --- &.804 & --- & 166 \\
\hline
\end{tabular}\end{center}

\medskip
The baryon conversion efficiency for low-mass
($10^{8.5}<(M_\text{Halo}/M_{\odot})< 10^{9.5}$) and high-mass
($10^{10.5}<(M_\text{Halo}/M_{\odot})$) halos, and total number of stellar
systems, for 
the No UV and FG UV backgrounds at various epochs. Statistical errors
are $\sim 20\%$ for the first six rows.  The last two rows, below the
line, are the $200^3$ 
simulations of halo A.
\end{table}

\begin{figure}
\includegraphics[width=\linewidth]{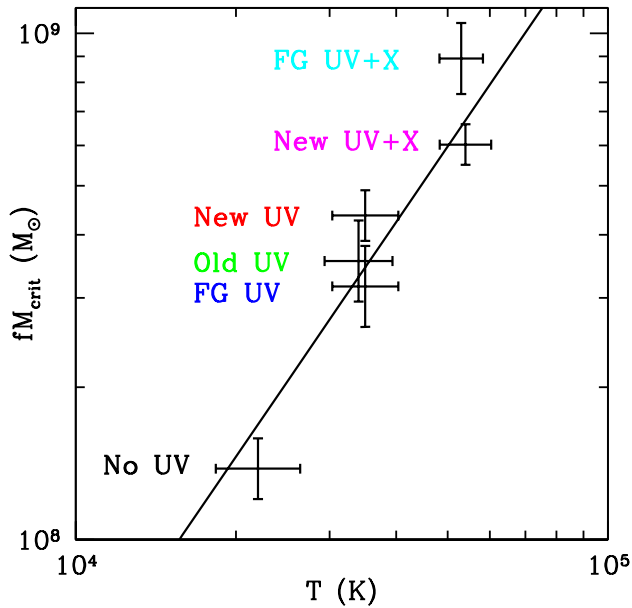}
\caption{The critical star-free halo mass at $z=0$ versus the mean gas
  temperature at $z=3$ for each background model, together with the
  best-fit power law.   Vertical and
  horizontal error bars represent 
  $1\sigma$ statistical errors among the 7 ICs.  We find $\log
M_{\text{crit}}\propto (1.54\pm 0.17) \log T$, in good agreement with
  the theoretical value of 1.5.}   
\label{fig:MT}
\end{figure}

Next, we test the validity of our physical argument by comparing
the mean gas temperature of the simulations at $z=3$ (when the small
systems are forming) to their critical
cutoff masses at $z=0$.  (In calculating the mean gas temperature we
do not separate dense galactic gas from the IGM; however at $z=3$ only
$\sim5\%$ of gas is at virial densities, so any contamination is less
than the variance of our ICs.)  Our theory predicts $M_{\text{crit}}\propto
T^{1.5}$.  Figure~\ref{fig:MT} shows the data from our six background
models together with the best-fit power law.  We find that $M_{\text{crit}}\propto T^{1.54\pm  0.17}$, in good agreement with
the theoretical value.  The goodness-of-fit statistic
$\chi^2/\text{d.f.}=1.8$ for this fit.  Notice that even in the No UV
case stellar feedback and dynamical heating are sufficient for the gas
to reach $2\times 10^4$~K by $z=3$ (the gas has reionized by this
epoch), but the UV heating increases the temperature by a factor of
1.6, and the X-rays by another factor of 1.5.  
  
Finally, we examine the level of numerical convergence in our
results.  We perform simulations of galaxy A with $200^3$ SPH
particles (i.e. twice the spatial resolution and eight times the mass
resolution) with the No UV and FG UV backgrounds (and including
feedback).  Due to computational constraints, the No~UV simulation was
run only to $z=2.5$, and the FG UV to $z=0.5$.  At $z=2.5$, the
Schechter $\alpha$ values for No UV and FG UV with $200^3$ resolution
are $-1.39$ and $-1.33$, respectively.  These values are more similar
than the $100^3$ runs, which have $\alpha=-1.45$ for No UV and $\alpha
= -1.07$ for FG UV at the same redshift.  The steepening slope for FG
UV as resolution 
increases is explained by smaller stellar systems which are
allowed by the higher peak densities associated with increased mass
resolution, and which merge away at late 
times to flatten the low-mass slope once again (whereas No~UV can keep
making new small systems).  Indeed, at $z=0.5$,
FG UV $200^3$ has $\alpha=-1.10$, compared to the $100^3$ run
$\alpha=-0.95$ (and No UV $100^3$ $\alpha=-1.57$).   Meanwhile, the No
UV $200^3$ simulation at $z=2.5$ still has a large number of small
stellar systems which have not yet formed stars, and whose gas is
disrupted by the background in the FG UV run.  In fact, 65\% of the
identified halos in No~UV $200^3$ at $z=2.5$ are in this category,
compared to only 39\% for the $100^3$ case.  However, the overall
baryon-conversion 
efficiency is not significantly different as a function of
resolution, at either $z=2.5$ ($<8\%$ difference) or $z=0.5$
($3\%$). The last two rows of Table~\ref{tab:hist} show the
breakdown by mass.  FG~UV $200^3$ has a much lower efficiency than
$100^3$ in the
low mass bin at $z=0.5$.  This is the result of 
 the same mechanism just discussed: the increased
resolution allows small halos to form and merge while remaining
star-free because of the UV background.  
Further, at $z=2.5$ FG~UV $200^3$ has a higher efficiency than
No~UV $200^3$. This is again the effect of proximity: while No~UV has
twice the number of halos in the low-mass bin as FG~UV, FG~UV's are
preferentially close to the central galaxy and thus have lower virial
masses and more efficient star formation, and increased numerical
resolution allows them to exist even 
closer without being lost to the halo finder.  (At the
25\textsuperscript{th} percentile of galaxies, FG UV has roughly a tenth the
efficiency of No UV in this mass bin at both resolutions.)   
Finally, the
$fM_\text{crit}$ statistic is not strongly affected by the increase in
resolution: at $z=2.5$, $\log fM_{\text{crit}}=(8.54,8.76)$ for
(No UV, FG UV), compared to $(8.47,8.85)$ for the respective $100^3$
runs.  
Thus we assert 
 that our results at $z=0$ would not be qualitatively
different with increased resolution, and quantitatively different by at
most 15\%.

\section{Discussion \& Conclusions}\label{sect:disc}  

In \citetalias{self09}, we examined the detailed effects on galaxy
properties of modest
changes to the ionizing radiation background (specifically the intensity
at high $z$).  Here we have simulated seven different galaxies with
their satellites, using six different models of the ionizing
background, and found that any UV background is
sufficient to replicate the observational results of the low end of
the galaxy mass spectrum, as parametrized by  Schechter $\alpha$.
The details of the spectrum and redshift dependence of the background
seem to have no 
significant effect.  We find that the number of small ($< 10^9
M_{\odot}$ in stars) systems at $z=0$ correlates with the critical
star-free halo mass, which in turn correlates well with the mean gas
temperature at $z=3$.  Further, the value of our critical mass
agrees with other recent studies of small galaxies.  These simulation
results are consistent with our 
theoretical picture, where only halos with an escape velocity greater
than the sound speed of the gas can retain their gas long enough for
it to cool and form stars. 

Thus we find the addition of an X-ray background causes
additional suppression of low-mass halos (by further increasing the
mean temperature of the gas), making the low-mass spectrum too
flat compared with observations.  It is possible that this is the
result of optical depth/self-shielding effects not being included in
the code, thus making the UV background more efficient at
heating/ionization than it is in reality. Optical depth in general
decreases with harder radiation, so the optically-thin assumption is
valid for X-rays.  However, the small halos
which we study here 
have the lowest densities and thus are the least effective
self-shielders: \citet{sawl09}
found that systems below the threshold mass of $10^9 M_{\odot}$ had
their gas densities sufficiently reduced by feedback and that
self-shielding of the UV background was not important. Our resolution
study weakly indicates that $\alpha$ may decrease with increasing
resolution, and a future paper (Hambrick et al., 2010, in prep.) finds
that including AGN thermal feedback causes $\alpha$ to decrease
modestly as well, presumably by driving gas back out of the central
galaxy to form stars in satellites.   This is certainly an
area deserving further study.  

We also study the baryon-conversion efficiencies of small halos, and
find that for small halos in the neighborhood of massive ones, and
find that the background is at least equally important as SN feedback
for lowering the efficiency to observed values, in contrast to
isolated dwarfs where feedback is the dominant mechanism. 

In sum, reproducing the observed numbers of low-mass galaxies in
hydrodynamic simulations is no
problem at all,  being
adequately resolved by the heating mechanisms already included in
standard hydro codes: feedback and the ionizing background.  

JPO was supported by NSF grant AST 07-07505 and NASA grant
NNX08AH31G. DCH thanks Ludwig Oser for invaluable debugging
assistance. TN and PHJ acknowledge support by the 
DFG cluster of excellence `Origin and Structure of the Universe'.

\end{document}